\documentclass[aps,pre,superscriptaddress,twocolumn,showpacs]{revtex4}
\usepackage{times,xspace}
\usepackage{latexsym,amsbsy,amssymb,amsmath,bm}
\usepackage{epsfig}

\begin {document}
\title {Synchronization of Random Linear Maps}
\author{Adam Lipowski}
\affiliation{Department of Physics, University of Geneva, CH 1211
Geneva 4, Switzerland}
\affiliation{Faculty of Physics, A.~Mickiewicz University,
61-614 Pozna\'{n}, Poland}
\author{Ioana Bena}
\affiliation{Department of Physics, University of Geneva, CH 1211
Geneva 4, Switzerland}
\author{Michel Droz}
\affiliation{Department of Physics, University of Geneva, CH 1211
Geneva 4, Switzerland}
\author{Antonio L.~Ferreira}
\affiliation{Departamento de Fisica, Universidade de Aveiro, 3810-193
 Aveiro, Portugal}
 %%%%%%%%%%%%%%%%%%%%%%%%%%%%%%%%%%%%%%%%%%%%%%%%%%%%%%%%%%%%%%%%%%%%%%%%%%%%%%%
\pacs{05.45.Xt}
\begin {abstract}
We study synchronization of random one-dimensional linear maps for which the
Lyapunov exponent can be calculated exactly.
Certain aspects of the dynamics of these maps are explained using their relation
with a random walk.
We confirm that the Lyapunov exponent changes sign at the complete
synchronization transition.
We also consider partial synchronization of nonidentical
systems.
It turns out that the way partial synchronization manifests depends on the
type of differences (in Lyapunov exponent or in contraction points) between
the systems.
The crossover from partial synchronization to complete synchronization is also
examined.
\end{abstract}
\maketitle
%%%%%%%%%%%%%%%%%%%%%%%%%%%%%%%%%%%%%%%%%%%%%%%%%%%%%%%%%%%%%%%%%%%%%%%%%%%%%
\section{Introduction}

Synchronization of chaotic systems is a subject of current
intensive study~\cite{PREP}.
To a large extent this is  due to its various applications, ranging from
laser dynamics~\cite{EXP} to electronic circuits~\cite{EXTENDED},
chemical and biological systems~\cite{BIOL}, secure
communications~\cite{SECURE}, etc.
But there is also a pure theoretical interest in this phenomenon, that
is related perhaps
to its counterintuitive nature: how it is possible that chaotic (i.e.,
by definition unpredictable) systems can
be synchronized and thus brought under some `control'.
And, even more puzzling, noise can play the role of the synchronizing factor.
Indeed, early reports~\cite{MARITAN} that sufficiently strong noise can
completely synchronize two identical chaotic systems were initially met with
scepticism and attributed
to finite precision of computations~\cite{PIKO} or to biased
noise~\cite{HERZEL}.
However, more recent examples show this effect even for unbiased
noise~\cite{TORAL}.

Since real systems are typically nonidentical,  complete synchronization
is difficult to achieve.
It is an interesting problem to examine whether noise can induce some sort
of `weaker synchronization' in nonidentical but relatively similar systems.
Recent works do show the existence of such partial
synchronization~\cite{ROSENBLUM,ZKURTHS}.

One  important problem of the theoretical and numerical studies of
synchronization is how to detect it.
This problem is essentially solved for the complete synchronization of
two identical systems that are described by variables $x$ and $x'$,
respectively.
In this case, for the synchronized state the difference
$|x-x'|$ equals zero, while it remains positive in the unsynchronized state.
Moreover, the transition between these two states is accompanied by the change
of the sign of the largest Lyapunov exponent, that becomes negative in
the synchronized state.

However, for the partial synchronization this problem is much
more subtle.
In this case, the two systems are not identical and the difference
$|x-x'|$ always remains positive.
It has already been noted for some models with continuous dyna\-mics
that partial synchronization manifests through changes
in the probability distribution of the `phase
difference'~\cite{ZKURTHS}.
In addition to that, in the partially synchronized phase the so-called
zero Lyapunov exponent becomes negative~\cite{ROSENBLUM,ZKURTHS}.

Studies of synchronization rely,  to a large extent,
on numerical calculations.
Precise estimations of Lyapunov exponents or probability distributions
(invariant measures)
constitute very often demanding computational problems.
To further test the already accumulated knowledge on
synchronization, it would be desirable to find models where at least
some of these properties could be computed analytically.

In the present paper we examine synchronization of random one-dimensional
linear maps.
For such maps one can easily find the exact Lyapunov exponent and
locate the point where it changes sign.
Numerical calculations for two identical systems confirm that this is
also the point where a complete synchronization transition takes place.
We briefly report on a correspondence between such maps and a random walk
process, that allows for a simple interpretation of
the initial stages of the evolution of the maps.

We also examine the partial synchronization of
nonidentical maps.
It is seen that the way partial synchronization manifests depends on the type of
difference between the nonidentical systems.
In a certain case, the difference in the location of contraction points of
the maps is imprinted in the probability
distribution at the partial synchronization transition.
When the difference between two subsystems $\delta$ tends to vanish,
partial
synchronization approaches the complete synchronization.
Due to the exact knowledge of the complete synchronization point,
one can examine some
details of this crossover.
In particular, it is shown that for $\delta\rightarrow 0$, vanishing
of the synchronization error is very slow $\sim (-1/{\rm log}_{10}\;\delta)$.

\section{Random 2-maps}
First, let us consider the simplest example of a random linear
map
\begin{equation}
x_{n+1}=f_i(x_n), \ i=0,1\ {\rm and} \ n=0,1,\ldots
\label{e1}
\end{equation}
where at each time step $n$ one of the maps $f_0$ or $f_1$ is applied
with a probability $p$ and $(1-p)$, respectively.
The maps are defined as $f_0(x)=a\,x\;{\rm mod}(1)$, and $f_1(x)=b\,x$,
where $0<x<1$ and $a>1,\ 0<b<1$.
Related models have already been examined in the context of
on-off intermittency~\cite{HEAGY,YANG}, advection of particles by chaotic
flows~\cite{ROMEIRAS}, and others~\cite{IRVIN,VULPIANI}.
Some aspects of synchronization were also studied for piecewise 
linear random maps,
but both the Lyapunov exponent and the location of the
synchronization transition were
determined only numerically~\cite{KOCAREV}.

For the map (\ref{e1}) it is elementary to calculate exactly its
Lyapunov exponent $\lambda$ defined as
\begin{equation}
\lambda=\underset{N\rightarrow\infty}{{\rm lim}}\frac{1}{N}
\sum_{n=0}^{N-1} {\rm log}_{10}\Big\arrowvert\frac{dx_{n+1}}{dx_n}\Big\arrowvert\;.
\label{e2}
\end{equation}
Indeed, since both $f_0$ and $f_1$ have constant derivatives, one immediately
obtains
\begin{equation}
\lambda=p\;{\rm log}_{10}\,a+(1-p)\;{\rm log}_{10}\,b\;.
\label{e3}
\end{equation}
It follows that $\lambda$ changes sign at
\begin{equation}
p=p_c=\frac{-{\rm log}_{10}\,b}{{\rm log}_{10}\,(a/b)}\;.
\label{e4}
\end{equation}
It is easy to understand this result.
For $p>p_c$ the expanding (chaotic) map $f_0$ prevails over the contracting
map $f_1$ and the overall behaviour is chaotic with $\lambda>0$.
The opposite situation takes place for $p<p_c$ and the map (\ref{e1}) contracts
to $x=0$.
At $p=p_c$ and in its close vicinity the map (\ref{e1})
exhibits intermittent bursts of
activity~\cite{HEAGY,YANG}, but we will not focus on such a behaviour
in the present paper.

To study synchronization one can make two runs $\{x_n\}$ and $\{x_n'\}$
of iterations of map
(\ref{e1}) starting each time from (slightly) different initial conditions
$x_0$ and $x_0'$ but with the same realization of noise, i.e., with the same
sequence of maps $f_i$.
Then one measures the synchronization error $w_n$ defined as
\begin{equation}
w_n=\langle|x_n-x_n'|\rangle \,,
\label{e6}
\end{equation}
where $\langle...\rangle$ represents the mean over the realizations of the
noise.
Moreover, we introduce the steady-state average
$w= \mbox{lim}_{n\rightarrow \infty}w_n$.
In the synchronized state $w=0$ while it is positive in the unsynchronized
state.
Our numerical results for $a=1/b=3/2$ (not presented here)
show that $w$ vanishes at $p=1/2$, i.e., the point where the Lyapunov exponent
changes sign.

However, due to the fact that it contracts to $x=0$ for $p<p_c$,
the map (\ref{e1}) is
not quite suitable to study synchronization of chaotic systems.
For such a purpose it would be desirable to have a map with a somehow
more complex
behaviour in the regime with a negative Lyapunov exponent.

Nevertheless, the simplicity of the map (\ref{e1})  allows us to get some
additional insight into its dynamics.
Let us fix the initial points
of our maps as, e.g., $x_0<x_0' \ll 1$.
As we shall see, such a choice results in a certain transient regime that 
can be understood using an analogy with a random walk process.

\begin{figure}
\centerline{\epsfxsize=9cm\epsfbox{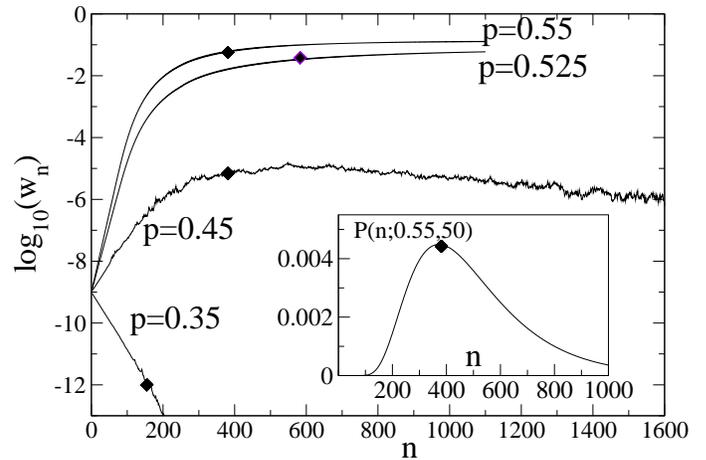}}
\caption{
The logarithm of the synchronization error $w_n$ as a function of
time $n$ for the map (\ref{e1}) with $a=1/b=3/2$, for different values of $p$.
The mean in $w_n$ was taken over $10^{10}$
realizations of the
random trajectories, and the initial points were $x_0=x_0'/2=10^{-9}$.
The diamonds locate the maximum $\tau_P$ of the corresponding probability
$P(n;p,y_0=50)$.
The inset shows a typical such first-passage probability $P(n;p=0.55,y_0=50)$
for the corresponding random walker to go from $y_0$ to $0$ in
$n$ steps. 
}
\label{fig1}
\end{figure}

Numerical evaluation of $w_n$ shows that it exhibits
three types of behavior as a function of time $n$ (Fig. \ref{fig1}).\\
(a) For $p>p_c$ (positive Lyapunov exponent)  after an
initial exponential increase, $w_n$ saturates and
acquires a constant nonzero value.\\
(b) For $p\lesssim p_c$, $w_n$ decreases after an initial
exponential increase.
The asymptotic exponential decay is consistent with the (negative) Lyapunov
exponent (\ref{e3}).\\
(c) For even smaller values of $p$ the synchronization error $w_n$ decreases
exponentially already from the beginning.

To understand the initial behaviour of $w_n$, let us recall that
the iteration of $x_n$ and $x_n'$ starts from very small values and
for a certain number of iterations the $\mbox{mod}(1)$ part of the map
(\ref{e1}) does not play any role.
Consequently, one has
\begin{equation}
w_n\,=\,\langle\alpha\rangle^{n}\,w_0\,,
\label{initial}
\end{equation}
where
\begin{equation}
\langle\alpha\rangle=p\,a+(1-p)\,b\,.
\end{equation}
Correspondingly, the initial decrease or increase of $w_n$ depends whether
$p$ is smaller or greater than $(1-b)/(a-b)$.

After a certain time, the ${\rm mod}(1)$ part of the map
comes into play and the initial behaviour of $w_n$
(\ref{initial}) is replaced by a different one.
Namely, for $p>p_c$, $w_n$ saturates at a nonzero value and it decays
exponentially for $p<p_c$.
We shall return to this point at the end of this section.

To estimate the time scale $\tau$ when the inital behaviour
(\ref{initial}) changes, we relate our map to a random walk process.
For simplicity, we consider the case $b=1/a$.
In this case there is a one-to-one correspondence between the stochastic
variable $x_n$ and
$y_n=-\mbox{log}_{a}\,(x_n)$.
Multiplication of $x_n$ by $a$ or $1/a$ corresponds to
the decrease or increase of $y_n$ by unity.
Consequently, $y_n$ is nothing else but the
position of a random walker on a lattice of unit spacing with transition
probabilities $p$ to the
left and $(1-p)$ to the right.
The correspondence with the random walk holds as long as the mod(1) part
of the map is not applied, i.e., the walker does not cross ${\rm log}_a(1)=0$.
The above random-walker problem has two characteristic time scales
connected to the
first-passage process~\cite{FELLER} from its initial position to 0, 
and one can expect that one of them is related to
$\tau$ [which is, recall, the time scale on which the initial behavior of the
map (\ref{initial}) is altered by the $\mbox{mod}(1)$ part of the map]. 
(i) First, there is the mean first passage time $\tau_M$. 
But for $p\leq p_c$, $\tau_M$ is infinite - contray to $\tau$, 
and therefore one cannot use $\tau_M$ as a measure for $\tau$.
(ii) Secondly, there is the time moment $\tau_P$ 
when the probability that the walker
hits $0$ for the first time is maximal. 
The probability distribution of the first hit is  known to 
be~\cite{FELLER}:
\begin{eqnarray}
P(n;p,y_0)&=&\displaystyle\frac{y_0}{n}\,
\left(
\begin{array}{c}
n \\
\displaystyle\frac{n-y_0}{2}
\end{array}
\right)
\displaystyle\,p^{\frac{(n+y_0)}{2}}
(1-p)^{\frac{(n-y_0)}{2}}\;,\nonumber\\
&&
\label{prob}
\end{eqnarray}
where $y_0$ is the initial position of the walker and $n$ is the number of
steps; the binomial coefficient is to be interpreted as zero
if $({n-y_0})/{2}$
is not an integer in the interval $[0,\,n]$.
Our numerical simulations show (see Fig. \ref{fig1})
that the value of $n=\tau_P$ for which $P(n;p,y_0)$ in 
Eq.~(\ref{prob}) attains a maximum offers a reasonable estimation of $\tau$.
One cannot expect to get a more precise estimate of $\tau$, 
since the change of the initial behavior (\ref{initial}) of the map 
towards the asymptotic one
is a gradual process, that involves 
all the trajectories of the equivalent random walker 
that hit 0 (for the first time) before as well as
after $\tau_P$.

Let us notice that the bias of the random walk is related with the sign of the
Lyapunov exponent and therefore with the asymptotic behaviour of our map.
Indeed, for $p<p_c=1/2$ the random walk is biased toward $+\infty$, that
translates into an exponential decay of $w_n$.
Our numerical simulations for the longer time regime suggest that this decay is
governed by the Lyapunov exponent $\lambda$ (\ref{e3}).
Recall that $\lambda$ is known to govern the evolution of the typical
value of $|x_n-x_n'|$ (see, e.g.,~\cite{VULPIANI}).
Thus in the long-time regime the mean $w_n$ and the typical value of
$|x_n-x_n'|$ behave identically.
On the other hand, they are clearly different in early-time regime.
It means that in this regime $w_n$ is strongly influenced by rare events, i.e.,
unlikely excursions of the random walker against the bias.
For $p>p_c=1/2$ the random walk is biased toward 0 and, since the map is
bounded, $w_n$ saturates at a positive value.

\section{Random 3-maps}
As already mentioned, the map (\ref{e1}) has a trivial behaviour for
$p<p_c$ and is not suitable to study synchronization of chaotic systems.
In this context the following  3-map version is more interesting:
\begin{equation}
x_{n+1}=f_i(x_n), \ i=0,1,2,
\label{e5}
\end{equation}
where $f_0(x)=a\,x\;{\rm mod}(1)$, $f_1(x)=b\,x$, and $f_2(x)=b\,x+(1-b)$.
The maps $f_0$, $f_1$, and $f_2$ are applied at random with probabilities $p$,
${(1-p)}/{2}$, and ${(1-p)}/{2}$, respectively.
It is easy to show that for such a random map Eqs.~(\ref{e3})-(\ref{e4}) still
hold.

\begin{figure}
\centerline{\epsfxsize=9cm
\epsfbox{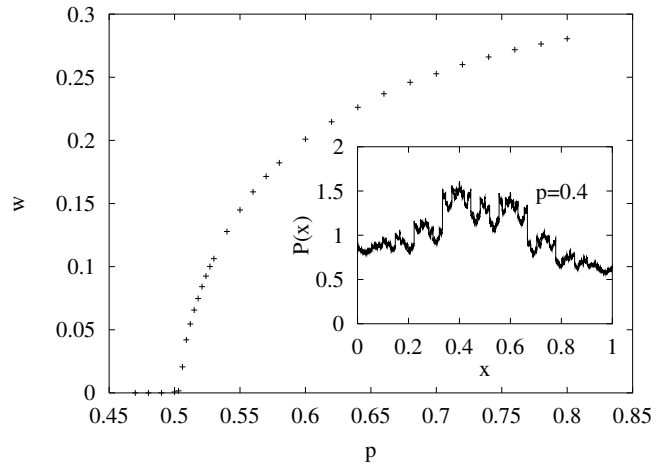}
}
%\figspace
\caption{The synchronization error $w$ as a function of $p$ calculated for the map
(\ref{e5}) with $a=1/b=3/2$.
The unnormalized probability distribution $P(x)$ of visiting a point $x$
for $p=0.4$ (synchronized phase) is shown in the inset.
The data were collected in $10^4$ bins.
Complex probability distributions are known to appear also
for some other random linear
maps~\cite{IRVIN}.}
\label{steadysame}
\end{figure}

Numerical evaluation of the synchronization error $w=\mbox{lim}_{n\rightarrow
\infty} w_n$ for the map (\ref{e5}) with $a=1/b=3/2$,
based on $N=10^8$ iterations, is shown in Fig.~\ref{steadysame}.
For $p>p_c=1/2$ the Lyapunov exponent $\lambda$ is positive and the system is
not synchronized ($w>0$).
For $p<1/2$ we have $\lambda<0$ and the system synchronizes ($w=0$).
But this time the behaviour for $p<1/2$ is much more complex.
The two maps $f_1$ and $f_2$ are still contracting ones, but to two different points (0 and 1).
Since they are applied randomly, the system, while remaining synchronized,
irregularly wanders throughout the whole interval $(0,1)$.
The probability density $P(x)$ of visiting a given point $x$, that is shown in
the inset of Fig.~\ref{steadysame}, confirms such a behaviour.
%%%%%-----------------------------------------------
%%%--------------------------------------------------

Random linear maps can be also used to study {\it partial synchronization}
that might
occur for two dynamical systems that are not identical, although
relatively similar.
Generally, one  considers the following pair of maps
\begin{equation}
x_{n+1}=f_i(x_n),\  x_{n+1}'=f_i'(x_n'),\ i=0,1,2\,,
\label{twomaps}
\end{equation}
for which  $\{f_0,\,f_0'\}$, $\{f_1,\,f_1'\}$, and $\{f_2,\,f_2'\}$
are applied at random with probabilities $p$,
${(1-p)}/{2}$, and ${(1-p)}/{2}$, respectively.
To complete the definition we have to specify the functions $f_i$ and $f_i'$,
$i=0,1,2$.
We present below the results for two particular choices. 
These ones correspond,
respectively, to a perturbation of the 
Lyapunov spectrum  - case (A) - and of
the attractor  - case (B).\\

\begin{figure}
\centerline{\epsfxsize=9cm
\epsfbox{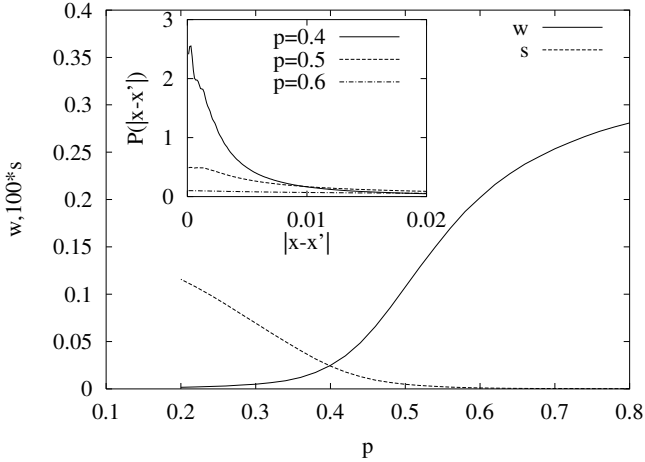}
}
%\figspace
\caption{
The synchronization error $w$ and the probability $s$ that $|x-x'|<10^{-4}$
as a function of $p$ for map (\ref{twomaps})-(\ref{e7}).
The unnormalized probability distribution $P(|x-x'|)$ is shown in
the inset.
}
\label{steadyslope}
\end{figure}

\noindent Case (A): We choose
\begin{equation}
f_0'(x)=a'\,x\;{\rm mod}(1),\ f_1'(x)=b'\,x,\ f_2'(x)=b'\,x+(1-b')
\label{e7}
\end{equation}
where $a'=3/2+\delta$, $b'=1/a'$ and $\delta=10^{-3}$.
Functions $f_0$, $f_1$, and $f_2$ are defined as for the map (\ref{e5})
with $a=1/b=3/2$.
For such a choice, the Lyapunov exponent of the map in Eq.~(\ref{e7})
is a linear function of $p$
that also changes sign at $p=p_c=1/2$ albeit with a different slope.
Assuming that a partial synchronization transition is also governed by the
Lyapunov exponent, we might expect that such a transition, if it exists,
takes place at $p=1/2$.
For the pair of maps (\ref{twomaps}), $w_n$ is defined as in Eq.~(\ref{e6}) but
$x_n$ and $x_n'$ evolve according to Eq.~(\ref{twomaps}).
Numerical calculation of $w={\rm lim}_{n\rightarrow\infty}w_n$ for $N=10^8$ 
iterations shows
(Fig.~\ref{steadyslope}) that it remains positive and smooth around this value.
However, there is a more subtle change in the system at or around $p=1/2$.
Indeed, the probability distribution of $P(|x-x'|)$ shows a pronounced peak
in the $p<1/2$ domain (inset of Fig.~\ref{steadyslope}).
This is yet another example that shows that partial synchronization
manifests  through a change in the probability distribution~\cite{ZKURTHS}.
However, $P(|x-x'|)$ was calculated for $p=0.4,\ 0.5$, and $0.6$, i.e., values
that are relatively far from each other.
For smaller differences between the values of $p$ the corresponding curves are
getting similar and
it is not clear to us whether there is a qualitative change in the
probability distribution $P(|x-x'|)$ that could locate precisely the partial
synchronization transition.
If not, it could mean that in this case either there is not a
well-defined partial
synchronization transition, or we are not looking at the right quantity to
detect it.

For  further comparison, we also calculated the probability $s$ that the
difference $|x-x'|$ remains smaller (if initially so)
than a given value $\varepsilon=10^{-4}$
during the iterations of the  system (\ref{twomaps}).
One finds that $s$ is a monotonous function of $p$
(Fig.~\ref{steadyslope}).\vspace{0.5cm}\\
%%%%%-----------------------------------------------
%%%--------------------------------------------------
\noindent Case (B): We choose
\begin{eqnarray}
&&f_0'(x)=a'\,x\;{\rm mod}(1),\; f_1'(x)=b'\,x,\nonumber\\
&& f_2'(x)=b'\,x+(1-b')-b''\;,
\label{e8}
\end{eqnarray}
where $a'=3/2$, $b'=2/3$ and $b''=10^{-3}$.
Functions $f_0$, $f_1$, and $f_2$ are defined as in case (A).
For such a choice the Lyapunov exponents of both systems are the same
and they change sign at $p=p_c=1/2$.
The only difference is that the function $f_2(x)$ has a contracting point at
$1$, while $f_2'(x')$ at a slightly smaller value $x'=0.997$.

\begin{figure}
\centerline{\epsfxsize=9cm
\epsfbox{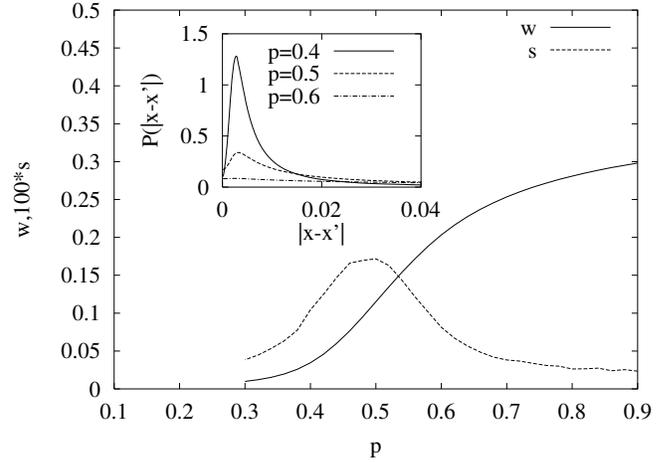}
}
%\figspace
\caption{
The synchronization error $w$ and the probability $s$ that $|x-x'|<10^{-4}$
as a function of $p$ for map (\ref{twomaps})-(\ref{e8}).
The unnormalized probability distribution $P(|x-x'|)$ is shown in
the inset.
}
\label{steady}
\end{figure}

\begin{figure}
\centerline{\epsfxsize=9cm
\epsfbox{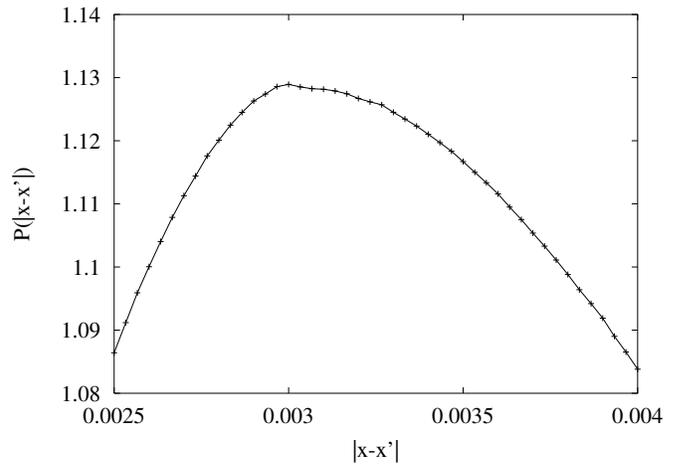}
}
%\figspace
\caption{
The unnormalized probability distribution $P(|x-x'|)$ calculated
for map (\ref{twomaps})-(\ref{e8}) at $p=0.5$.
The maximum at $|x-x'|=0.003$ is due to the shift in the location of
contracting points.
Off the synchronization transition ($p=0.5$) the location of the maximum
changes.
}
\label{dist05}
\end{figure}

In this case the synchronization error $w$ behaves similarly to case (A)
(Fig.~\ref{steady}).
However, the probability distribution of $P(|x-x'|)$ has a little bit different
shape.
In particular, even at $p=1/2$ it has a certain peak which increases with
decreasing $p$.
The maximum of the peak at $p=1/2$ is  most likely located at
$|x-x'|=0.003$ (Fig.~\ref{dist05}) and this is related with the
difference of the contracting points of the functions $f_2$ ($x=1$) and
$f_2'$ ($x'=0.997$).
We checked that off the $p=1/2$ point the maximum shifts away from
$|x-x'|=0.003$.
Moreover, the probability $s$ that the difference $|x-x'|$ is smaller than
$\varepsilon=10^{-4}$ shows a maximum at $p=1/2$ (Fig.~\ref{steady}).
%%%%%-----------------------------------------------
%%%-------------------------------------------------
%%%%%-----------------------------------------------
%%%--------------------------------------------------

Comparison of partial synchronization in  cases (A) and (B) revealed
important differences between them.
In  case (A), where we perturbed the Lyapunov exponent,
the probability distribution
$P(|x-x'|)$ develops a peak around/at $|x-x'|=0$
in the partially synchronized state that presumably exists for $p<1/2$.
Such a feature is similar to those already reported in the
literature~\cite{ZKURTHS}.
However, it is not clear to us whether in this case such a change can be
characterized more quantitatively so that a well-defined transition point
exists.
A different behaviour takes place in the case (B) with perturbed
contracting points.
In this case the peak of the probability distribution
is shifted by a value that at the partial synchronization
transition $p=1/2$ (as deduced from the vanishing of the Lyapunov exponent)
curiously matches the shift of the contracting points.
In addition to that, the probability of being in a state with a small
difference $|x-x'|$ has a maximum at $p=1/2$ -- in  drastic contrast
with  case (A).

For simple maps as those examined in the present paper one has a complete
knowledge of the Lyapunov exponents and fixed points.
This is usually not the case for more complicated dynamical systems like
Lorentz or R\"ossler equations.
Perturbing some parameters of these equations one usually modifies both the
Lyapunov spectrum and the attractors.
It is henceforth possible that partial synchronization in such
systems combines some
features of both cases (A) and (B) we discussed.

%%%%%-----------------------------------------------
\begin{figure}
\centerline{\epsfxsize=9cm
\epsfbox{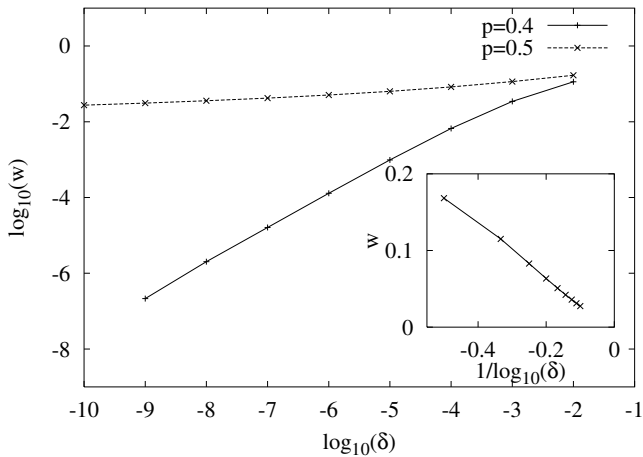}
}
%\figspace
\caption{
The synchronization error $w$ as a function of the mismatch $\delta$ for maps
(\ref{e7}).
The inset shows the data only at the partial
synchronization transition $p=0.5$.
}
\label{diff}
\end{figure}
%%%--------------------------------------------------
When the difference between nonidentical systems vanishes, partial
synchronization is replaced by  complete synchronization.
We shall now briefly examine such a situation.
In particular we consider two nonidentical systems as those
described in the case (A) above,
with varying difference $\delta$.
We calculated the synchronization error $w$ as a function of $\delta$ for
$p=0.4$ and $p=1/2$.
In both cases we expect that $w$ vanishes for $\delta\rightarrow 0$.
Figure~\ref{diff} confirms such an expectation, but it also reveals that for
$p=1/2$ the convergence to zero is much slower than for $p=0.4$.
While for $p=0.4$ the synchronization error seems to vanish as $w\sim \delta$,
the inset in Fig.~\ref{diff} suggests that at the synchronization transition
the vanishing is most likely logarithmic $w\sim ({-1}/{{\rm log}_{10}\,\delta)}$.

\section{Conclusions}
In the present paper we examined synchronization of one-dimensional
random maps.
Our results confirm that a complete synchronization transition
coincides with the change of sign of the Lyapunov exponent.
We also showed that the way the partial synchronization manifests depends on
the type of difference between the two nonidentical systems.
It would be desirable to explain the origin of the logarithmically
slow crossover of the partial synchronization to the complete synchronization.
It would be also interesting to explain why the location of the maximum of the
probability distribution at the partial synchronization transition discussed in
case (B) above matches exactly the shift of the contracting points.

\acknowledgments{This work was partially supported by the Swiss
National Science Foundation
and the project OFES 00-0578 "COSYC OF SENS".
A.~L.~F. is grateful for the financial support from
the Sapiens project POCTI/33141/99.}
%%%%%%%%%%%%%%%%%%%%%%%%%%%%%%%%%%%%%%%%%%%%%%%%%%%%%%%%%%%%%%%%%%%%%%%%%%%%%%
%%%%%%%%%%%%%%%%%%%%%%%%%%%%%%%%%%%%%%%%%%%%%%%%%%%%%%%%%%%%%%%%%%%%%%%%%%%%%%

%%%%%%%%%%%%%%%%%%%%%%%%%%%%%%%%%%%%%%%%%%%%%%%%%%%%%%%%%%%%%%%%%%%%%%%%%%%%%%%
%%%%%%%%%%%%%%%%%%%%%%%%%%%%%%%%%%%%%%%%%%%%%%%%%%%%%%%%%%%%%%%%%%%%%%%%%%%%%%%
\end {document}